# THE RARITY OF DNA PROFILES[1]

By Bruce S. Weir

*University of Washington*

It is now widely accepted that forensic DNA profiles are rare, so it was a surprise to some people that different people represented in offender databases are being found to have the same profile. In the first place this is just an illustration of the birthday problem, but a deeper analysis must take into account dependencies among profiles caused by family or population membership.

**1. Introduction.** In the 20 years since the introduction of DNA profiles for forensic identification there has developed a wide-spread belief that it is unlikely two people will share the same profile. Assuming at least 10 alleles or 55 genotypes at each locus, a 13-locus system in common use allows for at least $10^{21}$ different profiles, which far exceeds the total number of people in the world. It is difficult to attach a meaningful estimate to the probability that a person chosen at random would have a particular profile, but a good first step is to assume independence of all (26) alleles in a profile to arrive at an estimate that "reaches a figure altogether beyond the range of the imagination" in the language Galton (1892) used to describe probabilities for fingerprints. Given such arguments, what is to be made of recent findings that the profiles of two people in a database of offender profiles either match or come very close to matching? Is there a need to re-think the understanding that profiles are rare?

There are forensic, statistical and genetic aspects to discussions of profile rarity. The key forensic issue centers on the comparison of two profiles, often one from a crime-scene sample and one from a suspect. The relevant calculations must recognize the existence of two profiles rather than focusing on only one of them. The statistical aspects are addressed initially by the "Birthday Problem." The probability that a person chosen randomly has a particular birthday is 1/365, ignoring leap-year complications, but there

Received March 2007; revised August 2007.
[1]Supported in part by NIH Grant GM 075091.

is over 50% probability that two people in a group of 23 people share a birthday. This result recognizes that the number of *pairs* of people, 253, is much greater than the number of people, 23, and that the particular shared birthday is not specified. The finding of DNA profile matching in an Arizonan database of 65,000 profiles [Troyer, Gilroy and Koeneman (2001)] becomes less surprising when it is recognized that there are over two billion possible pairs of profiles in that database. The genetic aspects rest on the shared evolutionary history of humans. The very fact that the population is finite means that any two people have shared ancestors and the resulting dependencies increase the probability of profile matching.

**2. Forensic issues.** The interpretation of DNA forensic evidence $E$ requires the probabilities of that evidence under alternative hypotheses, referred to here as $H_p$ and $H_d$ for the case where they represent the views of prosecution and defense in a criminal trial. A simple scenario is when the profile $G_C$ of a crime-scene stain matches that, $G_S$, of a suspect. The hypotheses may be as follows:

$H_p$:  the suspect is the source of the crime-scene stain.

$H_d$:  the suspect is not the source of the crime-scene stain.

A quantity of interest to those charged with making a decision is the posterior odds of the prosecution hypothesis after the finding of matching DNA profiles:

$$\text{Posterior odds} = \frac{\Pr(H_p|E)}{\Pr(H_d|E)}.$$

From Bayes' theorem,

$$\frac{\Pr(H_p|E)}{\Pr(H_d|E)} = \frac{\Pr(E|H_p)}{\Pr(E|H_d)} \times \frac{\Pr(H_p)}{\Pr(H_d)},$$

$$\text{Posterior odds} = \text{LR} \times \text{Prior odds}$$

and it is the likelihood ratio LR that is estimated by forensic scientists. In paternity disputes this quantity is called the paternity index. Those who equate $\Pr(E|H_p)$ and $\Pr(H_p|E)$, as in "The odds were billions to one that the blood found at the scene was not O.J.s" [Anonymous (1997)], are said to have committed the "Prosecutor's Fallacy" [Thompson and Schumann (1987)].

The likelihood ratio for a single-contributor DNA profile can be expressed as

$$\text{LR} = \frac{\Pr(G_S, G_C|H_p)}{\Pr(G_S, G_C|H_d)}$$

$$= \frac{\Pr(G_S|G_C, H_p)}{\Pr(G_S|G_C, H_d)} \frac{\Pr(G_C|H_p)}{\Pr(G_C|H_d)}$$

$$= \frac{1}{\Pr(G_S|G_C, H_d)}$$

by recognizing that the crime-scene stain profile does not depend on the alternative hypotheses and that the two profiles must match under the prosecution hypothesis. Among the many advantages of adopting this approach to comparing competing hypotheses is the clarification that it is match probabilities $\Pr(G_S|G_C)$ for profiles from two people that are relevant rather than profile probabilities $\Pr(G_S)$. In the discussion of matching profiles in a database, $G_C$ and $G_S$ can refer to the profiles from different people and the issue is whether or not matching is unlikely.

**3. Statistical issues.** Diaconis and Mosteller (1989) discussed basic statistical techniques for studying coincidences and stated the law of truly large numbers: "With a large enough sample, any outrageous thing is likely to happen." Can we attach probabilities for very unlikely events to occur? In the forensic context, Kingston (1965) addressed match probabilities long before the advent of DNA profiling. If a particular item of evidence has a probability $P$, then he assumed that the unknown number $x$ of occurrences of the profile in a large population of $N$ people is Poisson with parameter $\lambda = NP$. Suppose a person with the particular profile commits a crime, leaves evidence with that profile at the scene, and then rejoins the population. A person with the profile is subsequently found in the population and a simple model says that the probability that this suspect is the perpetrator is $1/x$. Although $x$ is not known, it must be at least one, so the probability that the correct person has been identified is the expected value of $1/x$ given that $x \geq 1$. Those people who would equate $x$ to its expected value $\lambda$ and then assign equal probabilities to all $\lambda$ people are said to have committed the "Defense Attorney's Fallacy" [Thompson and Schumann (1987)].

Balding and Donnelly (1995), referring to Eggleston (1983) and Lenth (1986), pointed out that Kingston's conditioning on at least one individual having the profile is not the same as the correct conditioning, that a specific individual (the suspect) has the profile. They gave a general treatment of this "island problem" and then Balding (1999) followed with a discussion of uniqueness of DNA profiles. He started with the event that a person (the perpetrator) sampled at random from a population of size $(N + 1)$ has a particular profile. The remaining people in the population each have independent probability $P$ of having the same profile. A second person (the suspect) is drawn from the population and may be the same person as the first (event $G$). The second person is found to have the same profile as the first (event $E$). If $U$ is the event that the suspect has the profile and that no-one else in the population has the profile, then

$$\Pr(U|E) = \Pr(U|G, E)\Pr(G|E).$$

Now $\Pr(G|E) = \Pr(E|G)\Pr(G)/[\Pr(E|G)\Pr(G) + \Pr(E|\bar{G})\Pr(\bar{G})]$ by Bayes' theorem, and $\Pr(E|G) = 1$, $\Pr(E|\bar{G}) = P$, $\Pr(G) = 1/(N+1)$. Moreover, for independent profiles, $\Pr(U|G, E) = (1-P)^N$ so that $\Pr(U|E) > 1 - 2\lambda$. For the USA, with a population of about $3 \times 10^8$, a profile with a probability of $10^{-10}$ would give $\lambda = 0.03$ and the probability that the correct person has been identified of at least 0.94. This is not as dramatic a number as the original $10^{-10}$.

The birthday problem has to do with multiple occurrences of *any* profile, not a particular profile as treated by Kingston and Balding. Mosteller [(1962)](#) refers to the latter as the "birthmate problem." The probability that at least two of a sample of $n$ people have the same unspecified birthday (or DNA profile), in the case where every birthday (or profile) has the same probability $P$, is

$$\Pr(\text{At least one match}) = 1 - \Pr(\text{No matches})$$

$$= 1 - \{1(1-P)(1-2P)\cdots[1-(n-1)P]\}$$

$$\approx 1 - \prod_{i=0}^{n-1} e^{-iP} \approx 1 - e^{-n^2 P/2}$$

For the USA example of $P = 10^{-10}$, the chance of some profile being replicated in the population of $N = 3 \times 10^8$ is essentially 100%. The Arizona Department of Public Safety [Troyer, Gilroy and Koeneman [(2001)](#)] reported a nine-locus match in a database of 65,493 for a profile that had an estimated probability of 1 in $7.54 \times 10^8$. Using that probability, the chance of finding two matching profiles in the database would be about 94%, so the finding is not unexpected. DNA profiles do not have equal or independent probabilities, however, so these calculations are approximate at best.

**4. Genetic issues.** DNA profiles are genetic entities and, as such, are shaped by the evolutionary history of a population. Whereas it is sufficient to take samples from a population to provide descriptive statistics of that particular population, predictions of matching probabilities that recognize evolutionary events are necessarily expectations over replicate populations. There is no reason to believe that a particular population has properties that are at expectation.

As a simple example, consider the estimation of profile probabilities at a single locus **A**. If a sample of $n$ genotypes provides estimates $\tilde{p}_i$ for the frequencies $p_i$ of alleles $A_i$, then genotypic frequency estimates are $\tilde{p}_i^2$ for homozygotes $A_iA_i$ and $2\tilde{p}_i\tilde{p}_j$ for heterozygotes $A_iA_j$ under the assumption of random mating within the population. Taking expectations of these estimates, over repeated samples from the same population and over replicates of the sampled population, provides

$$\mathcal{E}(\tilde{p}_i^2) = p_i^2 + p_i(1-p_i)\left[\theta + \frac{1 + (2n-1)\theta}{2n}\right],$$

$$\mathcal{E}(2\tilde{p}_i\tilde{p}_j) = 2p_ip_j + 2p_ip_j\left[\theta + \frac{1 + (2n-1)\theta}{2n}\right]$$

[Weir (1996)] to introduce the population coancestry coefficient $\theta$ which measures the relationship between pairs of alleles within a population relative to the relationship of alleles between populations. To illustrate the meaning of "relative to" consider a fanciful example of a large community of people, all of whom are first cousins to each other. If these people pair at random, their children will form a population in which genotypic frequencies are products of allele frequencies. A child's two alleles, one from each parent, are independent. From the perspective of an observer outside the community, however, the allele pairs within the community appear to be dependent, with $\theta = 1/16$. This value of $\theta$ is needed to predict genotypic frequencies for the community children on the basis of population-wide allele frequencies.

For large sample sizes, the expected genotypic frequencies reduce to the parametric values $p_i^2 + p_i(1-p_i)\theta$ and $2p_ip_j(1-\theta)$. The sample allele frequencies $\tilde{p}_i$ are unbiased for the parametric values $p_i$ and $\theta$ is serving to provide the variance of the sample values—in particular, $p_i(1-p_i)\theta$ is the variance over populations of allele frequencies within one population. In the situation where alleles are selectively neutral, it is convenient to regard $\theta$ as the probability that a random pair of alleles in the same population are identical by descent, ibd, meaning that they have both descended from the same ancestral allele. Identity by descent is also an expectation over replicate populations.

The probabilities of pairs of genotypes require measures of relationship analogous to $\theta$ but for up to four alleles. Two individuals that are both homozygous $A_iA_i$ for the same allelic type, for example, may carry two, three, four or two pairs of alleles that are ibd. For the class of evolutionary models where there is stationarity under the opposing forces of mutation introducing genetic variation and genetic drift causing variation to be lost, and allelic exchangeability, these higher-order ibd probabilities may all be expressed in terms of $\theta$. The distribution of allele frequencies over replicate populations is Dirichlet for this class of models and a very useful consequence is that the probability of drawing an allele of type $A_i$ from a population given that $n_i$ of the previous $n$ alleles drawn were of that type is $[n_i\theta + (1-\theta)p_i]/[1 + (n-1)\theta]$ [Balding and Nichols (1997)]. This provides, for example, the probability of two members of the same population being homozygotes $A_iA_i$:

$$\Pr(A_iA_i, A_iA_i) = \frac{p_i[\theta + (1-\theta)p_i][2\theta + (1-\theta)p_i][3\theta + (1-\theta)p_i]}{(1+\theta)(1+2\theta)}.$$

From this and similar expressions for other genotypes, it is possible to predict the probability that two members of a population will match, that is, have the same two alleles at a locus [Weir (2004)],

$$\begin{aligned}
P_2 &= \sum_i \Pr(A_iA_i, A_iA_i) + \sum_i \sum_{j \neq i} \Pr(A_iA_j, A_iA_j) \\
&= \sum_i \Pr(A_iA_iA_iA_i) + 2 \sum_i \sum_{j \neq i} \Pr(A_iA_iA_jA_j) \\
&= \frac{1}{D}[6\theta^3 + \theta^2(1-\theta)(2 + 9S_2) \\
&\qquad + 2\theta(1-\theta)^2(2S_2 + S_3) + (1-\theta)^3(2S_2^2 - S_4)].
\end{aligned}$$

The first line specifies the genotypes, the second shows the corresponding sets of alleles, and the third shows the value from the Dirichlet assumption. Random mating is assumed for the second line. The third line employs the notation $S_k = \sum_i p_i^k, k = 2, 3, 4$, and $D = (1+\theta)(1+2\theta)$.

Partial matches occur when two individuals share one allele at a locus, rather than the two required for a match. As Diaconis and Mosteller (1989) said: "We often find 'near' coincidences surprising." The probability that two individuals partially match is

$$\begin{aligned}
P_1 &= 2 \sum_i \sum_{j \neq i} \Pr(A_iA_i, A_iA_j) + \sum_i \sum_{j \neq i} \sum_{k \neq i,j} \Pr(A_iA_j, A_iA_k) \\
&= 4 \sum_i \sum_{j \neq i} \Pr(A_iA_iA_iA_j) + 4 \sum_i \sum_{j \neq i} \sum_{k \neq i,j} \Pr(A_iA_iA_jA_k) \\
&= \frac{1}{D}[8\theta^2(1-\theta)(1 - S_2) + 4\theta(1-\theta)^2(1 - S_3) \\
&\qquad + 4(1-\theta)^3(S_2 - S_3 - S_2^2 + S_4)],
\end{aligned}$$

with the same meaning for the three rows as for $P_2$. Finally, for two individuals to mismatch, that is, have no alleles in common,

$$\begin{aligned}
P_0 &= \sum_i \sum_{j \neq i} \Pr(A_iA_i, A_jA_j) + 2 \sum_i \sum_{j \neq i} \sum_{k \neq i,j} \Pr(A_iA_i, A_jA_k) \\
&\quad + \sum_i \sum_{j \neq i} \sum_{k \neq i,j} \sum_{l \neq i,j,k} \Pr(A_iA_j, A_kA_l)
\end{aligned}$$

$$= \sum_i \sum_{j \neq i} \Pr(A_i A_i A_j A_j) + 2 \sum_i \sum_{j \neq i} \sum_{k \neq i,j} \Pr(A_i A_i A_j A_k)$$

$$+ \sum_i \sum_{j \neq i} \sum_{k \neq i,j} \sum_{l \neq i,j,k} \Pr(A_i A_j A_k A_l)$$

$$= \frac{1}{D}[\theta^2(1-\theta)(1-S_2) + 2\theta(1-\theta)^2(1-2S_2+S_3)$$

$$+ (1-\theta)^3(1-4S_2+4S_3+2S_2^2-3S_4)].$$

Values of $P_2$ are shown in Table 1 for 13 commonly-used forensic loci, using Caucasian allele frequencies reported by Budowle and Moretti (1999) and various values of $\theta$. Assuming independence of these loci, the full 13-locus match probabilities are the products of the 13 separate values and these products are also shown in Table 1. The probabilities of finding at least one matching pair among 65,493 individuals are given in Table 1, along with the sample size needed to give a 50% probability of at least one match. The column headed "Actual" shows the proportion of pairs of profiles that match at each locus in the very small sample of 203 Caucasian profiles reported by the FBI [Budowle and Moretti (1999)].

The finding of Troyer, Gilroy and Koeneman (2001) was for a pair of profiles that matched at nine loci, partially matched at three loci and mismatched at one locus. It is shown in Table 2 that, in fact, 163 such pairs of individuals are expected when loci are assumed to be independent and $\theta = 0.03$. This value of $\theta$ has been suggested as a very conservative value to use for forensic calculations [National Research Council (1996)], and Table 1 shows that value makes all 13 predicted match probabilities greater than FBI observed values. It would be of interest to examine the dataset of Troyer, Gilroy and Koeneman (2001) to see the level of agreement between observed and expected numbers of matches and partial matches. Weir (2004) was able to examine an Australian dataset of 15,000 profiles and showed (Table 4) very good agreement when $\theta$ was set to 0.001. The agreement was not as good when $\theta$ was set to zero. Table 3 shows observed and expected numbers of match/partial match combinations for the Caucasian data of Budowle and Moretti (1999). The sample size is too small to have more than six loci with matches and is really too small to allow strong conclusions about the role of $\theta$ to be made. This example shows good overall agreement between observed and expected values for $\theta = 0$. Examination of actual offender datasets is needed.

It is clear, however, that instances of matching and partially matching profiles are not unexpected in offender databases.

**5. Effect of relatives.** The previous results accommodated the effects of shared evolutionary history on the probabilities that two individuals have

TABLE 1

*Probabilities that two unrelated noninbred[1] people match at common loci, based on allele frequencies reported by Budowle and Moretti (1999)*

| Locus | Actual[2] | $\theta$ | | | | |
|---|---|---|---|---|---|---|
| | | **0.000** | **0.001** | **0.005** | **0.010** | **0.030** |
| D3S1358 | 0.077 | 0.075 | 0.075 | 0.077 | 0.079 | 0.089 |
| vWA | 0.063 | 0.062 | 0.063 | 0.065 | 0.067 | 0.077 |
| FGA | 0.036 | 0.036 | 0.036 | 0.038 | 0.040 | 0.048 |
| D8S1179 | 0.063 | 0.067 | 0.068 | 0.070 | 0.072 | 0.083 |
| D21S11 | 0.036 | 0.038 | 0.038 | 0.040 | 0.042 | 0.051 |
| D18S51 | 0.027 | 0.028 | 0.029 | 0.030 | 0.032 | 0.040 |
| D5S818 | 0.163 | 0.158 | 0.159 | 0.161 | 0.164 | 0.175 |
| D13S317 | 0.076 | 0.085 | 0.085 | 0.088 | 0.090 | 0.101 |
| D7S820 | 0.062 | 0.065 | 0.066 | 0.068 | 0.070 | 0.080 |
| CSF1PO | 0.122 | 0.118 | 0.119 | 0.121 | 0.123 | 0.134 |
| TPOX | 0.206 | 0.195 | 0.195 | 0.198 | 0.202 | 0.216 |
| THO1 | 0.074 | 0.081 | 0.082 | 0.084 | 0.086 | 0.096 |
| D16S539 | 0.086 | 0.089 | 0.089 | 0.091 | 0.094 | 0.105 |
| All loci | | $2 \times 10^{-15}$ | $2 \times 10^{-15}$ | $3 \times 10^{-15}$ | $4 \times 10^{-15}$ | $2 \times 10^{-14}$ |
| Prob.[3] | | 0.000,004 | 0.000,004 | 0.000,006 | 0.000,009 | 0.000,050 |
| Sample size[4] | | 28 million | 27 million | 22 million | 18 million | 7.7 million |

[1] Apart from evolutionary-driven inbreeding and relatedness.

[2] Observed proportion of matches in data of Budowle and Moretti (1999).

[3] Probability of at least one matching pair among 65,493 individuals.

[4] Sample size to give 50% probability of at least one match.

TABLE 2

*Expected numbers of pairs of matching or partially matching profiles in a sample of size 65,493 profiles when at least six of 13 loci match if $\theta = 0.03$*

| Number of matching loci | Number of partially matching loci | | | | | | | |
|---|---|---|---|---|---|---|---|---|
| | **0** | **1** | **2** | **3** | **4** | **5** | **6** | **7** |
| 6 | 4,059 | 37,707 | 148,751 | 322,963 | 416,733 | 319,532 | 134,784 | 24,125 |
| 7 | 980 | 7,659 | 24,714 | 42,129 | 40,005 | 20,061 | 4,150 | |
| 8 | 171 | 1,091 | 2,764 | 3,467 | 2,153 | 530 | | |
| 9 | 21 | 106 | 198 | 163 | 50 | | | |
| 10 | 2 | 7 | 8 | 3 | | | | |
| 11 | 0 | 0 | 0 | | | | | |
| 12 | 0 | 0 | | | | | | |
| 13 | 0 | | | | | | | |

the same genotype. These probabilities are increased if the individuals have a shared family history. Allowing for this degree of relatedness, but still

Table 3

*Observed and expected numbers of profiles with specified numbers of matching or partially loci when all 94 profiles in a dataset of Budowle and Moretti (1999) are compared to each other*

| No. of match- ing loci | θ | \multicolumn{14}{c}{Number of partially matching loci} | | | | | | | | | | | | | |
|---|---|---|---|---|---|---|---|---|---|---|---|---|---|---|---|
| | | **0** | **1** | **2** | **3** | **4** | **5** | **6** | **7** | **8** | **9** | **10** | **11** | **12** | **13** |
| 0 | Obs. | 0 | 3 | 18 | 92 | 249 | 624 | 1077 | 1363 | 1116 | 849 | 379 | 112 | 25 | 4 |
| | 0.000 | 0 | 2 | 19 | 90 | 293 | 672 | 1129 | 1403 | 1290 | 868 | 415 | 134 | 26 | 2 |
| | 0.001 | 0 | 2 | 18 | 88 | 286 | 661 | 1114 | 1391 | 1286 | 869 | 418 | 135 | 26 | 2 |
| | 0.010 | 0 | 2 | 14 | 70 | 236 | 566 | 992 | 1289 | 1241 | 875 | 439 | 148 | 30 | 3 |
| | 0.030 | 0 | 1 | 8 | 42 | 152 | 396 | 754 | 1065 | 1118 | 860 | 471 | 174 | 39 | 4 |
| 1 | Obs. | 0 | 12 | 48 | 203 | 574 | 1133 | 1516 | 1596 | 1206 | 602 | 193 | 43 | 3 | |
| | 0.000 | 0 | 7 | 50 | 212 | 600 | 1192 | 1704 | 1768 | 1320 | 692 | 242 | 51 | 5 | |
| | 0.001 | 0 | 7 | 49 | 208 | 592 | 1182 | 1698 | 1770 | 1328 | 700 | 246 | 52 | 5 | |
| | 0.010 | 0 | 5 | 40 | 178 | 527 | 1094 | 1637 | 1779 | 1393 | 767 | 282 | 62 | 6 | |
| | 0.030 | 0 | 3 | 26 | 125 | 401 | 905 | 1475 | 1749 | 1496 | 901 | 363 | 88 | 10 | |
| 2 | Obs. | 0 | 7 | 61 | 203 | 539 | 836 | 942 | 807 | 471 | 187 | 35 | 2 | | |
| | 0.000 | 1 | 9 | 56 | 210 | 514 | 871 | 1040 | 877 | 511 | 196 | 45 | 5 | | |
| | 0.001 | 1 | 9 | 56 | 208 | 512 | 872 | 1046 | 886 | 519 | 200 | 46 | 5 | | |
| | 0.010 | 1 | 8 | 50 | 193 | 494 | 875 | 1096 | 969 | 593 | 239 | 57 | 6 | | |
| | 0.030 | 0 | 5 | 38 | 160 | 445 | 861 | 1178 | 1140 | 765 | 339 | 89 | 11 | | |
| 3 | Obs. | 0 | 6 | 33 | 124 | 215 | 320 | 259 | 196 | 92 | 16 | 1 | | | |
| | 0.000 | 1 | 7 | 36 | 116 | 243 | 344 | 334 | 220 | 94 | 23 | 3 | | | |
| | 0.001 | 1 | 6 | 36 | 116 | 244 | 348 | 339 | 224 | 96 | 24 | 3 | | | |
| | 0.010 | 0 | 6 | 35 | 117 | 256 | 380 | 387 | 268 | 120 | 32 | 4 | | | |
| | 0.030 | 0 | 5 | 31 | 115 | 275 | 447 | 499 | 379 | 187 | 54 | 7 | | | |
| 4 | Obs. | 1 | 5 | 17 | 29 | 54 | 82 | 67 | 16 | 6 | 0 | | | | |
| | 0.000 | 0 | 3 | 15 | 40 | 70 | 81 | 61 | 29 | 8 | 1 | | | | |
| | 0.001 | 0 | 3 | 15 | 40 | 71 | 82 | 63 | 30 | 8 | 1 | | | | |
| | 0.010 | 0 | 3 | 15 | 44 | 81 | 98 | 78 | 40 | 12 | 1 | | | | |
| | 0.030 | 0 | 3 | 16 | 52 | 105 | 139 | 122 | 68 | 22 | 3 | | | | |
| 5 | Obs. | 0 | 1 | 2 | 6 | 12 | 14 | 6 | 5 | 0 | | | | | |
| | 0.000 | 0 | 1 | 4 | 9 | 13 | 11 | 6 | 2 | 0 | | | | | |
| | 0.001 | 0 | 1 | 4 | 9 | 13 | 12 | 7 | 2 | 0 | | | | | |
| | 0.010 | 0 | 1 | 4 | 11 | 16 | 15 | 9 | 3 | 0 | | | | | |
| | 0.030 | 0 | 1 | 6 | 15 | 25 | 26 | 17 | 6 | 1 | | | | | |
| 6 | Obs. | 0 | 1 | 0 | 2 | 2 | 0 | 0 | 0 | | | | | | |
| | 0.000 | 0 | 0 | 1 | 1 | 1 | 1 | 0 | 0 | | | | | | |
| | 0.001 | 0 | 0 | 1 | 1 | 2 | 1 | 0 | 0 | | | | | | |
| | 0.010 | 0 | 0 | 1 | 2 | 2 | 1 | 1 | 0 | | | | | | |
| | 0.030 | 0 | 0 | 1 | 3 | 4 | 3 | 1 | 0 | | | | | | |

TABLE 4
*Identity probabilities for common family relationships*

| Relationship | $k_2$ | $k_1$ | $k_0$ |
|---|---|---|---|
| Identical twins | 1 | 0 | 0 |
| Full sibs | $\frac{1}{4}$ | $\frac{1}{2}$ | $\frac{1}{4}$ |
| Parent and child | 0 | 1 | 0 |
| Double first cousins | $\frac{1}{16}$ | $\frac{3}{8}$ | $\frac{9}{16}$ |
| Half sibs | 0 | $\frac{1}{2}$ | $\frac{1}{2}$ |
| Grandparent and grandchild | 0 | $\frac{1}{2}$ | $\frac{1}{2}$ |
| Uncle and nephew | 0 | $\frac{1}{2}$ | $\frac{1}{2}$ |
| First cousins | 0 | $\frac{1}{4}$ | $\frac{3}{4}$ |
| Unrelated | 0 | 0 | 1 |

assuming random mating within a population so there is no inbreeding, requires the probabilities $k_2, k_1, k_0$ that the individuals have received 2, 1 or 0 pairs of alleles identical by descent from their immediate family ancestors. Values for these probabilities for common relationships are shown in Table 4. Individuals that share two pairs of ibd alleles must have matching genotypes. Those that share one pair of alleles ibd may either match or partially match, and individuals with no ibd allele sharing may match, partially match or mismatch. Therefore, the probabilities that two individuals match, partially match or mismatch at one locus are

$$\mathrm{Pr(Match)} = k_2 + k_1\left[\sum_i \mathrm{Pr}(A_iA_iA_i) + \sum_i\sum_{j \neq i} \mathrm{Pr}(A_iA_jA_j)\right] + k_0P_2$$

$$= k_2 + k_1[\theta + (1-\theta)S_2] + k_0P_2,$$

$$\mathrm{Pr(Partial\ Match)} = k_1\left[2\sum_i\sum_{j \neq i}\mathrm{Pr}(A_iA_iA_j) + \sum_i\sum_{j \neq i}\sum_{k \neq i,j}\mathrm{Pr}(A_iA_jA_k)\right] + k_0P_1$$

$$= k_1(1-\theta)(1-S_2) + k_0P_1,$$

$$\mathrm{Pr(Mismatch)} = k_0P_0.$$

Equivalent results were given by Fung, Carracedo and Hu (2003). Numerical values for the matching probabilities for the 13-locus system described in Table 1 are shown in Table 5 for common relationships. Clearly, the probabilities increase with the degree of relationship.

Pairs of relatives with related common ancestors within their family are inbred, and the three ibd probabilities $k_2, k_1, k_0$ must be replaced by a more extensive set of nine probabilities $\Delta_i, i = 1, 2, \ldots, 9$, for the various patterns of ibd among all four alleles carried by the two relatives [Weir, Anderson and Hepler (2006)]. These are defined in Table 6, along with numerical values for the situation of full sibs whose parents are first cousins. The various matching probabilities become

$$\Pr(\text{Match}) = (\Delta_1 + \Delta_7) + (\Delta_2 + \Delta_3 + \Delta_5 + \Delta_8)[\theta + (1 - \theta)S_2],$$

$$+ \frac{1}{1 + \theta}(\Delta_4 + \Delta_6)[2\theta^2 + 3\theta(1 - \theta)S_2 + (1 - \theta)^2 S_3]$$

TABLE 5
*Matching probabilities for common family relationships (with $\theta = 0.03$)*

| Locus | Not related | First-cousins | Parent–child | Full-sibs |
|---|---|---|---|---|
| D3S1358 | 0.089 | 0.124 | 0.229 | 0.387 |
| vWA | 0.077 | 0.111 | 0.213 | 0.376 |
| FGA | 0.048 | 0.078 | 0.166 | 0.345 |
| D8S1179 | 0.083 | 0.119 | 0.227 | 0.384 |
| D21S11 | 0.051 | 0.081 | 0.172 | 0.349 |
| D18S51 | 0.040 | 0.068 | 0.150 | 0.335 |
| D5S818 | 0.175 | 0.216 | 0.339 | 0.463 |
| D13S317 | 0.101 | 0.139 | 0.252 | 0.401 |
| D7S820 | 0.080 | 0.115 | 0.219 | 0.379 |
| CSF1PO | 0.134 | 0.173 | 0.288 | 0.428 |
| TPOX | 0.216 | 0.261 | 0.397 | 0.503 |
| THO1 | 0.096 | 0.133 | 0.241 | 0.395 |
| D16S539 | 0.105 | 0.143 | 0.256 | 0.404 |
| Total | $2 \times 10^{-14}$ | $2 \times 10^{-12}$ | $6 \times 10^{-9}$ | $5 \times 10^{-6}$ |

TABLE 6
*Identity probabilities for inbred relatives carrying alleles $(a, b)$ and $(c, d)$, and values for example of siblings whose parents are first cousins*

| ibd alleles | Probability | Example[*] |
|---|---|---|
| $a$, $b$, $c$, $d$ | $\Delta_1$ | 1/64 |
| $a$, $b$ and $c$, $d$ | $\Delta_2$ | 0 |
| $a$, $b$, $c$ or $a$, $b$, $d$ | $\Delta_3$ | 2/64 |
| $a$, $b$ only | $\Delta_4$ | 1/64 |
| $a$, $c$, $d$ or $b$, $c$, $d$ | $\Delta_5$ | 2/64 |
| $c$, $d$ only | $\Delta_6$ | 1/64 |
| $(a, c$ and $b, d)$ or $(a, d$ and $b, c)$ | $\Delta_7$ | 15/64 |
| $a$, $c$ or $a$, $d$ or $b$, $c$ or $b$, $d$ | $\Delta_8$ | 30/64 |
| none | $\Delta_9$ | 12/64 |

[*]First cousin provides alleles $a$, $c$ to sibs, second cousin provides alleles $b$, $d$ to sibs.

$$+ \Delta_9 P_2,$$

$$\Pr(\text{PartialMatch}) = (\Delta_3 + \Delta_5 + \Delta_8)(1 - \theta)(1 - S_2)$$

$$+ \frac{2(1 - \theta)}{1 + \theta}(\Delta_4 + \Delta_6)[\theta + (1 - 2\theta)S_2 - (1 - \theta)S_3]$$

$$+ \Delta_9 P_1,$$

$$\Pr(\text{Mismatch}) = \Delta_2(1 - \theta)(1 - S_2)$$

$$+ \frac{1 - \theta}{1 + \theta}(\Delta_4 + \Delta_6)[1 - (2 - \theta)S_2 + (1 - \theta)^2 S_3] + \Delta_9 P_0.$$

Relatedness will increase the probability that two individuals will have matching or partially matching DNA profiles and it would not be surprising if very large offender databases had profiles from related people. It is difficult, however, to turn the question around and infer relatedness of people whose profiles have a high degree of matching. The current set of less than 20 STR loci is not enough to give good estimates of the degree of relatedness [Weir, Anderson and Hepler (2006)], and even unrelated people can have very similar profiles.

**6. Discussion.** DNA profiling has proven to be a powerful tool for human identification in forensic and other contexts. Different people, identical twins excepted, have different genetic constitutions and it is hoped that an examination of a small portion of these constitutions will allow for identification or differentiation. Current forensic DNA profiling techniques examine between 10 and 20 regions of the genome, representing of the order of $10^3$ of the $10^9$ nucleotides in the complete genome. Nevertheless, the probability that a randomly chosen person has a particular forensic profile can easily reach the small value of $10^{-10}$. Even when the forensic scientist is careful to present probabilities in the preferred format such as "the probability of a person having this profile given that we know the perpetrator has the profile," the numbers remain small and the evidence that a defendant also has that profile can be compelling.

Given the widespread belief that specific forensic profiles are rare, there has been some concern expressed at the finding of matching or nearly matching profiles in databases of less than 100,000. Such findings were predicted by Weir (2004), unaware that they had already been reported [Troyer, Gilroy and Koeneman (2001)] for the case of two profiles matching at nine of 13 loci. At the simplest level, the apparent discrepancy is merely an application of the birthday problem. If all DNA profiles have the same probability $P$, and if profiles are independent, then the probability of at least two instances of any profile in a set of $n$ profiles is approximately $1 - \exp(-n^2 P/2)$. This probability can be large even for small $P$ and it can be 50% when $n$ is of the order of $1/\sqrt{P}$. The widespread practice of collecting profiles from people suspected of, arrested for, or convicted of crimes has already led to the establishment of large databases: the National DNA Database (NDNAD) in the United Kingdom had over three million profiles in February 2006 and the Combined DNA Index System (CODIS) in the United States had over four million profiles in February 2007. These and other national databases are growing.

This note has looked a little more closely at the probability of finding matching profiles in a database. The first observation was that DNA profiles are genetic entities with evolutionary histories that impose dependencies among profiles. The formulation of dependencies was made for single loci, but there is empirical evidence [Weir (2004), Figure 1] that sufficiently large "correction" for dependencies within loci will also accommodate between-locus dependencies. This means taking sufficiently large values of the parameter $\theta$.

Incorporation of "$\theta$-corrections" for the case of unrelated individuals refers to the dependencies generated by the evolutionary process. These would not be detected from observations taken solely within a population, but they are necessary to enable predictions to be made. Predictions need to take variation among populations into account. Additional dependencies due to nonrandom mating, leading to within-population inbreeding, were considered by Ayres and Overall (1999).

**Acknowledgment.** Very helpful comments were made by an anonymous reviewer.

Department of Biostatistics
University of Washington
Seattle, Washington 98195
USA
E-mail: bsweir@u.washington.edu